\documentclass[11pt]{article}
\usepackage[a4paper,margin=1in]{geometry}
\usepackage{amsmath,amsfonts}
\usepackage{algorithmic}
\usepackage{algorithm}
\usepackage{array}
\usepackage[caption=false,font=normalsize,labelfont=sf,textfont=sf]{subfig}
\usepackage{textcomp}
\usepackage{stfloats}
\usepackage{url}
\usepackage{verbatim}
\usepackage{graphicx}
\usepackage{soul} 
\usepackage{xcolor}
\usepackage{authblk}
\usepackage[style=ieee, backend=biber]{biblatex}
\begin{document}

\title{Periodic OFDMA: A Low-PAPR Multiple Access Scheme for Uplink Communications in 5G and Beyond}

\author[1]{Gokce Hacioglu}
\author[1]{Serkan Vela}
\affil[1]{Karadeniz Technical University}

\date{}

\maketitle

\begin{abstract}
Multiple access techniques are essential for 5G, 6G, and beyond, where a large number of users need to share resources efficiently. Orthogonal frequency division multiple access (OFDMA) is a conventional technique for multiple access. However, a high peak-to-average power ratio (PAPR), which reduces energy efficiency, makes OFDMA impractical, especially in uplink transmissions. This paper presents a novel multiple access scheme, Periodic OFDMA (P-OFDMA), which demonstrates a lower peak-to-average power ratio and reduced computational complexity compared to conventional OFDMA. The proposed P-OFDMA method assigns subcarriers to users in a periodic pattern, allowing all users to utilize the entire frequency band, thereby improving frequency diversity while simplifying subcarrier allocation. Furthermore, two precoded variants, P-OFDMA-DCT and P-OFDMA-DFT, are introduced to achieve lower PAPR; notably, P-OFDMA-DFT provides both the lowest PAPR and lower computational complexity. Through comprehensive simulations, the performance of P-OFDMA is compared with conventional OFDMA and Single-Carrier FDMA (SC-FDMA). The results demonstrate that P-OFDMA-DFT consistently achieves the lowest PAPR, making it a highly power-efficient solution. The standard P-OFDMA scheme also shows significant improvements, notably outperforming SC-FDMA in PAPR for scenarios with a low number of subcarriers per user and offering lower computational complexity. In addition, by assigning each user to periodically spaced subcarriers distributed across the full frequency band, P-OFDMA provides enhanced frequency diversity and achieves better BER performance than SC-FDMA under high delay-spread conditions. Moreover, the proposed P-OFDMA and its precoded versions (P-OFDMA-DCT and P-OFDMA-DFT) require substantially lower transmitter-side processing than SC-FDMA—up to an eightfold reduction—which is particularly beneficial for low-complexity uplink devices. Although the receiver side requires higher computational effort, the total processing load considering all uplink users together with the receiver remains lower overall, leading to improved energy efficiency. Consequently, P-OFDMA and its precoded enhancements present powerful and energy-efficient solutions for uplink communications in 5G and beyond.
\end{abstract}


\section{INTRODUCTION}

Multiple access techniques are a cornerstone of modern wireless communication systems, serving as fundamental enablers for efficient spectrum utilization and user accommodation, particularly in the context of 5G and beyond networks. As wireless networks evolve to support a diverse array of applications---ranging from enhanced mobile broadband (eMBB) to massive machine-type communications (mMTC) and ultra-reliable low-latency communications (URLLC)---the demand for flexible and highly efficient multiple access schemes has intensified significantly \cite{3gpp2017}. Among the established technologies, Orthogonal Frequency Division Multiple Access (OFDMA) has emerged as a widely adopted solution due to its inherent ability to provide orthogonal resource allocation, robust performance in frequency-selective channels, and adaptability to various traffic patterns \cite{myung2006}.

The deployment of 5G networks has introduced unprecedented challenges in terms of user density and service heterogeneity. Massive Machine-Type Communications (mMTC), in particular, presents unique requirements characterized by sporadic traffic from a vast number of devices, each typically requiring low data rates \cite{dawy2017}. While effective for broadband applications, traditional OFDMA systems face critical limitations in mMTC scenarios where users are allocated a small number of subcarriers \cite{shahab2020grantfree}. In such cases, the conventional approach of assigning contiguous resource blocks may yield lower BER performance in frequency-selective channels due to reduced frequency diversity, while also suffering from high Peak-to-Average Power Ratio (PAPR) and lower energy efficiency \cite{han2005}.

The high PAPR challenge is especially pronounced in uplink transmissions, where power amplifier efficiency directly impacts the battery life of user devices. A high PAPR necessitates a significant power amplifier back-off, which reduces energy efficiency and limits achievable data rates \cite{jiang2008}. This issue is paramount for the battery-constrained devices typical in mMTC, where ensuring an extended operational lifetime is a primary design goal \cite{dawy2017}.

To address this, recent research has explored various PAPR reduction techniques, such as selective mapping (SLM) and partial transmit sequences (PTS) \cite{bauml1996}. However, these methods often introduce significant computational complexity, rendering them unsuitable for the resource-constrained devices found in mMTC applications. Conversely, while Single-Carrier Frequency Division Multiple Access (SC-FDMA) offers inherently lower PAPR, it does so at the cost of reduced frequency diversity and system flexibility \cite{falconer2002}. The conventional block-based subcarrier allocation in OFDMA, though simple to manage, fails to exploit the full frequency diversity of the system when user allocations are small, creating a clear need for more advanced strategies \cite{baig2010}.

In this context, we propose a novel multiple access scheme termed Periodic Orthogonal Frequency Division Multiple Access (P-OFDMA), specifically designed to overcome the limitations of conventional OFDMA in scenarios characterized by low subcarrier allocation per user. The proposed P-OFDMA method leverages a periodic, non-contiguous subcarrier allocation pattern that allows each user to access the entire frequency band, thereby maximizing frequency diversity while maintaining orthogonality. This approach is particularly well-suited for 5G mMTC applications, where numerous devices require sporadic access to small spectrum resources.

The key contributions of this work are as follows:

\begin{itemize}
    \item We propose P-OFDMA, a novel multiple access scheme with periodic subcarrier allocation that provides enhanced frequency diversity, lower computation complexity at transmitter side and has low PAPR.
    \item We introduce two precoded variants, P-OFDMA-DCT and P-OFDMA-DFT. Both achieve superior PAPR performance; moreover, P-OFDMA-DFT has the lowest transmitter-side computational complexity.
    \item We show that the proposed schemes provide improved energy efficiency, making them attractive and practical solutions for battery-constrained devices in massive IoT deployments.
\end{itemize}

The remainder of this paper is organized as follows. Section II presents the system model and provides a detailed description of the proposed P-OFDMA method. Section III presents a comprehensive performance evaluation, including PAPR analysis, bit error rate (BER) performance, power efficiency, and computational complexity comparisons. Finally, Section IV concludes the paper with a summary of key findings and potential directions for future research.

\section{System Model and P-OFDMA}\label{sys}

OFDMA is based on the principle of orthogonal frequency division multiplexing, where symbols mapped through a modulation scheme undergo an Inverse Fast Fourier Transform (IFFT) before transmission. To mitigate inter-block interference and transform linear convolution into circular convolution, a cyclic prefix is appended to the IFFT output before performing digital-to-analog conversion. The inverse discrete Fourier transform is expressed as:
\begin{align}
	\label{IFFT}
	s(n-\ell) &= \frac{1}{\sqrt{N}} \sum_{k=0}^{N-1} S(k) e^{j \frac{2\pi k n}{N}} e^{-j\frac{2\pi k \ell}{N}} \nonumber \\
	&= \frac{1}{\sqrt{N}} \sum_{k=0}^{N-1} S(k) W_{N}^{k(\ell-n)}
\end{align}

\noindent where $S(k)$ denotes the samples to be transmitted, which represent symbols mapped based on the modulation scheme employed. The number of symbols in one OFDM block is $N$, and $s(n-\ell)$ represents the $\ell$-sample shifted version of $s(n)$, where $\ell=0,1,2,\ldots,N-1$.

The vector representation of $M$ symbols in the frequency domain, transmitted by the $m^{\text{th}}$ user, is expressed as:

\begin{equation}
	\label{vector_periodic}
	\mathbf{S_{m,f}} = \begin{bmatrix}
		S_{m,0} & S_{m,1} & \cdots & S_{m,(M-1)}
	\end{bmatrix}^T
\end{equation}

\noindent The matrix $\mathbf{A}$ is constructed by repeating the $M \times M$ identity matrix $\mathbf{I}_M$ exactly $K = \frac{N}{M}$ times, producing a matrix of size $N \times M$. Here, $K$ denotes the number of P-OFDMA users.
\begin{equation}
	\label{A_matrix}
	\mathbf{A} = \begin{bmatrix} \mathbf{I_M} & \mathbf{I_M} & \cdots & \mathbf{I_M} \end{bmatrix}^T
\end{equation}

\noindent The elements of the vector $\mathbf{S_{m,N}}$ are formed by repeating a set of $M$ samples each $K$ times:
\begin{equation}
	\label{S}
	\mathbf{S_{m,N}} = \mathbf{A} \mathbf{S_{m,f}}
\end{equation}

The matrix representation of the shifted version of the IFFT, as shown in Equation \eqref{IFFT}, is given below. Here, $\mathbf{F_{\ell,N}^H}$ represents the $\ell$-sample shifted version of the $N$-point IFFT. Moreover, the FFT matrix, $\mathbf{F_{\ell,N}}$, is equivalent to the Hermitian transpose of $\mathbf{F_{\ell,N}^H}$, i.e., $\mathbf{F_{\ell,N}} = (\mathbf{F_{\ell,N}^H})^\mathbf{H}$.

\begin{equation}
	\label{matrixifft}
	\mathbf{F_{\ell,N}^H} = \frac{1}{\sqrt{N}}
	\begin{bmatrix}
		1 & W_N^{\ell} & \dots & W_N^{\ell(N-1)} \\
		1 & W_N^{(\ell-1)} & \dots & W_N^{(\ell-1)(N-1)} \\
		1 & W_N^{(\ell-2)} & \dots & W_N^{(\ell-2)(N-1)} \\
		\vdots & \vdots & \ddots & \vdots \\
		1 & W_N^{(\ell-N+1)} & \dots & W_N^{(\ell-N+1)(N-1)}
	\end{bmatrix}
\end{equation}

For samples that periodically repeat themselves with a period of $M$, the user with offset $\ell \in \{0,1,\ldots,K-1\}$ will have an IFFT output different from zero at the indices $\ell, K+\ell, 2K+\ell, \ldots, (M-1)K+\ell$.

It is assumed that there are $K$ users and a single base station in the area. Communication between the users and the base station is carried out using the proposed periodic OFDMA (P-OFDMA) scheme. The circular channel matrix between the $m^{\text{th}}$ user $u_m$ and the base station, after removing the cyclic prefix, is given by:
\begin{equation}
	\label{H0}
	\begin{aligned}
		\mathbf{H_{m}} &=
		\begin{bmatrix}
		h_{m,0} & \cdots & h_{m,\mu_m} & 0 & \cdots & 0 \\
		0 & \cdots & h_{m,\mu_{m-1}} & h_{m,\mu_m} & \cdots & 0 \\
		\vdots & \ddots & \ddots & \ddots & \ddots & \vdots \\
		0 & \cdots & h_{m,0} & h_{m,1} & \cdots & h_{m,\mu_m} \\
		h_{m,\mu_m} & \cdots & 0 & h_{m,0} & \cdots & h_{m,\mu_{m-1}} \\
		\vdots & \ddots & \ddots & \ddots & \ddots & \vdots \\
		h_{m,1} & \cdots & h_{m,\mu_m} & 0 & \cdots & h_{m,0}
	\end{bmatrix} \\
		&= \mathbf{F_{0,N}^H} \mathbf{\Lambda_m} \mathbf{F_{0,N}}
	\end{aligned}
\end{equation}

\noindent where $u_m$ is assumed to have $\mu_m + 1$ taps. $\mathbf{\Lambda_m}$ is an $N \times N$ diagonal matrix, and $\mathbf{F_{0,N}}$ is the discrete Fourier transform (DFT) matrix.

A diagonal matrix, denoted as $\mathbf{X_m}$, is defined as follows:
\begin{equation}
	\label{Xm}
	\mathbf{X_m} = \frac{1}{\sqrt{K}}\mathbf{F_{0,N}} \mathbf{F_{\ell_m,N}^H}
\end{equation}
\noindent where $\ell_m \in \{0,1,\ldots,K-1\}$ denotes the time-domain shift (offset) assigned to user $m$.

User $m$ will transmit the $\mathbf{S_m}$ vector for the P-OFDMA scheme, which is represented below:
\begin{equation}
	\label{Sm}
	\mathbf{S_m} = \mathbf{X_m} \mathbf{A} \mathbf{F_{0,M}^H} \mathbf{S_{m,f}}
\end{equation}
To further reduce the PAPR, precoding techniques can be applied to the modulated symbols prior to the P-OFDMA mapping. The precoded variants, namely P-OFDMA-DCT and P-OFDMA-DFT, are obtained by multiplying the symbol vector $S_{m,f}$ with an $M$-point DCT matrix ($C_{M}$) or an $M$-point DFT matrix ($F_{0,M}$), respectively. The transmitted signal for P-OFDMA-DCT is given by:

\begin{equation}
\mathbf{S_{m}^{(\text{DCT})}} = \mathbf{X_{m}} \mathbf{A} \mathbf{F_{0,M}^{H}} \mathbf{C_{M}} \mathbf{S_{m,f}}
\label{eq:pofdma_dct}
\end{equation}

\noindent where $\mathbf{C_{M}}$ denotes the $M$-point Discrete Cosine Transform matrix. Similarly, for P-OFDMA-DFT, the signal is expressed as:

\begin{equation}
\mathbf{S_{m}^{(\text{DFT})}} =\mathbf{X_{m} A F_{0,M}^{H} F_{0,M} S_{m,f}}
\label{eq:pofdma_dft_raw}
\end{equation}

\begin{figure*}[!t]
	\centering
	\includegraphics[width=0.80\linewidth]{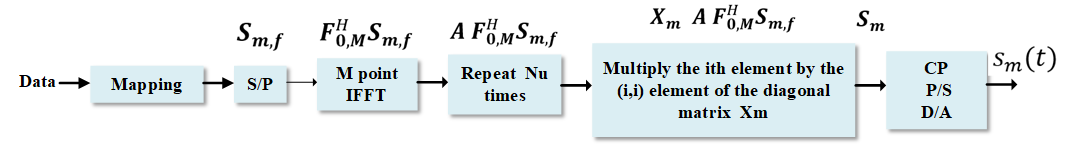}
	\caption{Uplink P-OFDMA Transmitter for User $m$}
	\label{verici}
\end{figure*}

\noindent It is important to note that in the P-OFDMA-DFT scheme, the DFT precoding matrix ($\mathbf{F_{0,M}}$) and the IFFT operation ($\mathbf{F_{0,M}^{H}}$) inherently cancel each other out, as $\mathbf{F_{0,M}^{H} F_{0,M}} = \mathbf{I_M}$. Consequently, the expression simplifies to a direct time-domain repetition and phase shift of the symbols, significantly reducing the computational complexity at the transmitter side:

\begin{equation}
\mathbf{S_{m}^{(\text{DFT})}} = \mathbf{X_{m} A I_M S_{m,f}} = \mathbf{X_{m} A S_{m,f}}
\label{eq:pofdma_dft_simp}
\end{equation}
\noindent Figure \ref{verici} presents the block diagram of the transmitter for user $u_m$.

After removing the cyclic prefix from the received vector $\mathbf{S_m}$ through an ISI channel, and neglecting AWGN noise, the received vector $\mathbf{R_m}$ for $u_m$ can be expressed as follows:
\begin{equation}
	\label{Rm}
	\begin{aligned}
		\mathbf{R_m} &= \mathbf{H_{m}} \mathbf{S_m}
		=\frac{1}{\sqrt{K}} \mathbf{F_{0,N}} \mathbf{\Lambda_m} \mathbf{F_{0,N}^H} \mathbf{X_m} \mathbf{A} \mathbf{F_{0,M}^H} \mathbf{S_{m,f}} \\
		&=\frac{1}{\sqrt{K}} \mathbf{F_{0,N}} \mathbf{\Lambda_m} \mathbf{F_{0,N}^H} \mathbf{F_{0,N}} \mathbf{F_{\ell_m,N}^H} \mathbf{A} \mathbf{F_{0,M}^H} \mathbf{S_{m,f}} \\
		&=\frac{1}{\sqrt{K}}\mathbf{F_{0,N}} \mathbf{\Lambda_m} \mathbf{F_{\ell_m,N}^H} \mathbf{A} \mathbf{F_{0,M}^H} \mathbf{S_{m,f}}
	\end{aligned}
\end{equation}

The receiver can multiply $\mathbf{R_m}$ with $(\mathbf{X_k} \mathbf{A} \mathbf{F_{0,M}^H})^H$ to obtain the transmitted symbols from user $m$. The result will be zero if $m \neq k$ in Equation \eqref{Rm2}.
\begin{equation}
	\label{Rm2}
	\begin{aligned}
		\mathbf{R_{m,k}}&=(\mathbf{X_k} \mathbf{A} \mathbf{F_{0,M}^H})^H \mathbf{R_m} \\
		&=\frac{1}{K} \mathbf{F_{0,M}} \mathbf{A^H} \mathbf{F_{\ell_k,N}} \mathbf{F_{0,N}^H} \mathbf{F_{0,N}} \mathbf{\Lambda_m} \mathbf{F_{\ell_m,N}^H} \mathbf{A} \mathbf{F_{0,M}^H} \mathbf{S_{m,f}} \\
		&=\frac{1}{K} \mathbf{F_{0,M}} \mathbf{A^H} \mathbf{F_{\ell_k,N}} \mathbf{\Lambda_m} \mathbf{F_{\ell_m,N}^H} \mathbf{A} \mathbf{F_{0,M}^H} \mathbf{S_{m,f}}
	\end{aligned}
\end{equation}

The term in Equation \eqref{Rm2} is denoted as $\mathbf{\tilde{\Lambda}_{m,k}}$ and is either a diagonal matrix or a zero matrix of size $M \times M$, as shown below:
\begin{equation}
	\label{lambda}
	\mathbf{\tilde{\Lambda}_{m,k}} =\frac{1}{K} \mathbf{F_{0,M}} \mathbf{A^H} \mathbf{F_{\ell_k,N}} \mathbf{\Lambda_m} \mathbf{F_{\ell_m,N}^H} \mathbf{A} \mathbf{F_{0,M}^H}
\end{equation}

The term $\mathbf{F_{0,M}} \mathbf{A^H} \mathbf{F_{\ell_k,N}}$ in the above equation has dimensions $M \times N$ and contains one nonzero element per row. The nonzero element in the $i^{\text{th}}$ row of $\mathbf{F_{0,M}} \mathbf{A^H} \mathbf{F_{\ell_k,N}}$ is located at $\text{mod}\left( N - \ell_k + K i, N \right)$. In the proposed P-OFDMA scheme, each user is assigned a unique offset (ID), i.e., the mapping $m \mapsto \ell_m$ is one-to-one. Therefore, if $\ell_m = \ell_k$ (equivalently, $m = k$), the matrix $\mathbf{\tilde{\Lambda}_{m,k}}$ will be a diagonal matrix, and the diagonal elements will be equal to the $\text{mod}\left( N - \ell_k + K i, N \right), \text{mod}\left( N - \ell_k + K i, N \right)$ element of $\mathbf{\Lambda_m}$, where $i = 0, 1, \hdots, M-1$. For example, with $N = 16$ and $K = 4$, there are $M = \frac{N}{K} = 4$ subcarriers. The channel coefficients for users $u_0$ and $u_1$ correspond to the subcarrier indices $0, 4, 8, 12$ and $15, 3, 7, 11$, respectively. Each user will possess $M$ channel coefficients, which are derived from the $N$ subcarriers and are separated by $K$. This means that all users will have channel coefficients that relate to both lower and higher frequency subcarriers, thereby exploiting full frequency diversity.

At the P-OFDMA receiver, the received signal $\mathbf{R_\Sigma}$, after cyclic prefix removal, represents a superposition of the transmitted vectors from all $K$ users. Each user's signal has traversed a distinct Inter-Symbol Interference (ISI) channel and is further corrupted by additive white Gaussian noise (AWGN), denoted by $\mathbf{W}$:
\begin{equation}
	\label{Rt}
	\mathbf{R_{\Sigma}} = \sum_{m=0}^{K-1} \mathbf{H_{m}}\mathbf{S_m}+ \mathbf{W}
\end{equation}

The uplink P-OFDMA receiver retrieves the symbols transmitted by $u_m$ as shown below. Since $\tilde{\Lambda}_{m,m}$ is a diagonal matrix, a single-tap equalizer is sufficient to eliminate inter-symbol interference:
\begin{equation}
	\label{smre}
	\mathbf{\tilde{{S}}_{m,f}}=
	\begin{bmatrix}
\tilde{S}_{m,0} \\
\tilde{S}_{m,1} \\
\vdots \\
\tilde{S}_{m,M-1}
\end{bmatrix}
=
\mathbf{\tilde{\Lambda}^{-1}_{m,m}} \mathbf{F_{0,M}} \mathbf{A^H X_m^H R_\Sigma}
\end{equation}
For the precoded P-OFDMA variants, the receiver structure requires an additional processing step to recover the original symbols. After the single-tap frequency domain equalization represented by the diagonal matrix $\mathbf{\tilde{\Lambda}^{-1}_{m,m}}$, the receiver must apply the inverse of the precoding matrix used at the transmitter. Since both DCT and DFT matrices are unitary, their inverses correspond to their transpose and Hermitian transpose, respectively. Therefore, the recovered symbol vectors for P-OFDMA-DCT and P-OFDMA-DFT are obtained by multiplying $\mathbf{\tilde{{S}}_{m,f}}$ with $\mathbf{C_M^T}$ and $\mathbf{F_{0,M}^H}$, as shown below:

\begin{equation}
\begin{aligned}
    \mathbf{\hat{S}_{m,f}^{(\text{DCT})}} &=\mathbf{C_M^T  \tilde{S}_{m,f}} \\
    \mathbf{\hat{S}_{m,f}^{(\text{DFT})}} &= \mathbf{F_{0,M}^H \tilde{S}_{m,f}}
\end{aligned}
\label{eq:precoded_receiver}
\end{equation}

\begin{figure*}[!t]
	\centering
	\includegraphics[width=0.70\linewidth]{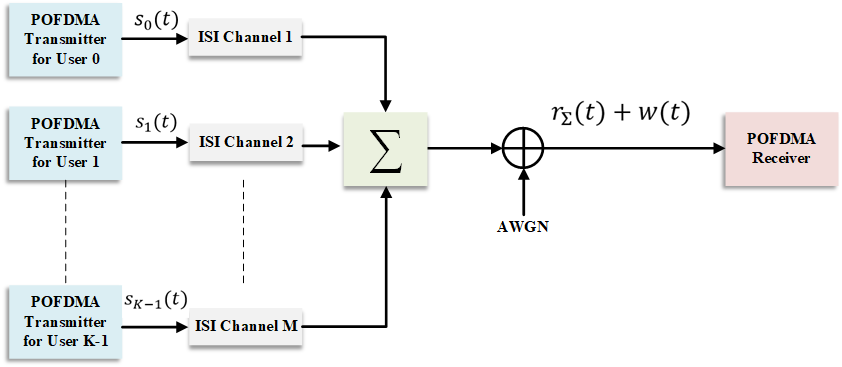}
	\caption{Uplink P-OFDMA System Model}
	\label{sistem}
\end{figure*}

\begin{figure*}[!t]
	\centering
	\includegraphics[width=0.8\linewidth]{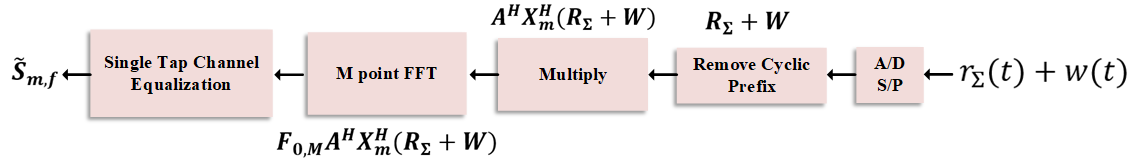}
	\caption{Uplink P-OFDMA Receiver }
	\label{alici}
\end{figure*}

\noindent A block diagram of the proposed P-OFDMA system is shown in Fig. \ref{sistem}, and the corresponding P-OFDMA receiver is depicted in Fig. \ref{alici}.

\textbf{Key Features of the Proposed P-OFDMA Scheme}

\textbf{Transmitter Side:}
\begin{itemize}
	\item The P-OFDMA scheme assumes a communication scenario where $K$ users interact with a base station.
	\item Each user transmits $M$ symbols, where $M = \frac{N}{K}$.
	\item Users are assigned unique identifiers, denoted by $m$, where $m = 0, 1, 2, \dots, K-1$.
	\item User $m$ ($u_m$) performs an $M$-point IFFT on the symbols to be transmitted, as defined in Equation \eqref{Sm}.
	\item The resulting IFFT output, $\mathbf{F_{0,M}^H} \mathbf{S_{m,f}}$, is either repeated $K$ times or multiplied by matrix $\mathbf{A}$.
	\item User $u_m$ then multiplies the repeated IFFT output by $\mathbf{X_m}$ to generate $\mathbf{S_m}$, which is then prepared for transmission.
	\item Before digital-to-analog conversion, a cyclic prefix is appended to $\mathbf{S_m}$ to mitigate inter-block interference.
\end{itemize}

\textbf{Receiver Side:}
\begin{itemize}
	\item To retrieve the transmitted symbols from $u_m$, the received samples $\mathbf{R_\Sigma}$ (after removing the cyclic prefix) are processed by multiplying them with $\mathbf{A^H}\mathbf{X_m^H}$.
	\item An $M$-point FFT is applied to the processed samples to transform them back into the frequency domain.
	\item Following the FFT operation, each $i^{\text{th}}$ sample ($i=0,1,\dots,M-1$) is equalized by dividing it by the corresponding channel coefficient from the diagonal of $\mathbf{\Lambda_m}$ at index $\text{mod}\left( N - \ell_m + K i, N \right)$.
\end{itemize}

\begin{table}[H]
\centering
\caption{COMPUTATIONAL COMPLEXITY ANALYSIS OF PROPOSED AND CONVENTIONAL SCHEMES}
\label{tab:complexity_formulas}
\renewcommand{\arraystretch}{1.5}
\resizebox{\textwidth}{!}{%
\begin{tabular}{|l|c|c|c|c|}
\hline
\multicolumn{1}{|c|}{} & \multicolumn{2}{c|}{\textbf{Transmitter (Tx)}} & \multicolumn{2}{c|}{\textbf{Receiver (Rx)}} \\ \cline{2-5}
\textbf{Method} & \textbf{Multiplications ($C_{\text{mult}}$)} & \textbf{Additions ($C_{\text{add}}$)} & \textbf{Multiplications ($C_{\text{mult}}$)} & \textbf{Additions ($C_{\text{add}}$)} \\ \hline
\textbf{OFDMA} & $\frac{N}{2} \log_2 N$ & $N \log_2 N$ & $\frac{N}{2} \log_2 N + N$ & $N \log_2 N$ \\ \hline
\textbf{SC-FDMA} & $\frac{N}{2} \log_2 N + \frac{M}{2} \log_2 M$ & $N \log_2 N + M \log_2 M$ & $\frac{N}{2} \log_2 N + K \left( \frac{M}{2} \log_2 M + M \right)$ & $N \log_2 N + K ( M \log_2 M )$ \\ \hline
\textbf{P-OFDMA} & $\frac{M}{2} \log_2 M + N$ & $M \log_2 M$ & $K \left( N + \frac{M}{2} \log_2 M + M \right)$ & $K ( M \log_2 M )$ \\ \hline
\textbf{P-OFDMA-DCT} & $M \log_2 M + N$ & $2M \log_2 M$ & $K \left( N + M \log_2 M + M \right)$ & $K ( 2M \log_2 M )$ \\ \hline
\textbf{P-OFDMA-DFT} & $N$ & $0$ & $K \left( N + M \log_2 M + M \right)$ & $K ( 2M \log_2 M )$ \\ \hline
\end{tabular}%
}
\end{table}

Each uplink P-OFDMA user experiences a distinct Inter-Symbol Interference (ISI) channel towards the P-OFDMA receiver, characterized by potentially different numbers of taps and varying channel coefficients.

\subsection{Computational Complexity}
The computational complexity for the transmitter (Tx) per user and the total computational load at the receiver (Rx) for processing all $K$ users are derived and listed in Table \ref{tab:complexity_formulas}. We assume a radix-2 FFT algorithm is employed.

The proposed scheme provides a fourfold reduction in the number of multiplications and fewer additions on the transmit side, which is critical given the typically lower processing capability of user equipment. Conversely, the receiver-side computational load is higher due to the user-specific phase operation involving $\mathbf{X}_m^{H}$ (Table \ref{tab:complexity_comparison}). However, because the base station can accommodate higher processing complexity, the relevant system-level metric is the total uplink processing load, i.e., the sum of all user-side transmit operations and the base-station receive operations.

Using Table \ref{tab:complexity_formulas}, the total operation load is first written as the sum of two components: (i) all user-side transmitter operations, obtained by multiplying per-user Tx complexity by the number of users $K$, and (ii) total receiver-side operations at the base station. Applying this aggregation for OFDMA and P-OFDMA yields total multiplication and addition expressions in terms of $(N,M,K)$. Then, by substituting the user-allocation relation $K=\frac{N}{M}$ into those total expressions and simplifying, the closed-form results in (18)--(21) are obtained:
\begin{alignat}{2}
& C_{\mathrm{mult,tot}}^{\mathrm{OFDMA}} &&= \left(\frac{N}{M}+1\right)\frac{N}{2}\log_2 N + N \\
& C_{\mathrm{add,tot}}^{\mathrm{OFDMA}}  &&= \left(\frac{N}{M}+1\right)N\log_2 N \\
& C_{\mathrm{mult,tot}}^{\mathrm{P\mbox{-}OFDMA}} &&= N\log_2 M + \frac{2N^2}{M} + N \\
& C_{\mathrm{add,tot}}^{\mathrm{P\mbox{-}OFDMA}}  &&= 2N\log_2 M
\end{alignat}

For a representative 1024-subcarrier case ($N=1024$) with $K=64$ ($M=16$), OFDMA requires $333{,}824$ total complex multiplications and $665{,}600$ total complex additions, whereas the proposed P-OFDMA requires $136{,}192$ total complex multiplications and $8{,}192$ total complex additions. Therefore, in this configuration, P-OFDMA reduces total multiplications by about $2.45\times$ and total additions by about $81.25\times$ compared with OFDMA.

\begin{table*}[!t]
\centering
\caption{COMPUTATIONAL LOAD (COMPLEX MULTIPLICATIONS ONLY) COMPARISON. TX VALUES ARE PER-USER, RX VALUES ARE TOTAL AT BS.}
\label{tab:complexity_comparison}
\resizebox{\textwidth}{!}{%
\begin{tabular}{|c|c|c||c|c||c|c||c|c||c|c||c|c|}
\hline
\multicolumn{1}{|c|}{} & \multicolumn{1}{c|}{} & \multicolumn{1}{c||}{} & \multicolumn{2}{c||}{\textbf{OFDMA}} & \multicolumn{2}{c||}{\textbf{SC-FDMA}} & \multicolumn{2}{c||}{\textbf{P-OFDMA}} & \multicolumn{2}{c||}{\textbf{P-OFDMA-DCT}} & \multicolumn{2}{c|}{\textbf{P-OFDMA-DFT}} \\
\cline{4-13}
\textbf{N} & \textbf{K} & \textbf{M} & \textbf{Tx (User)} & \textbf{Rx (Total)} & \textbf{Tx (User)} & \textbf{Rx (Total)} & \textbf{Tx (User)} & \textbf{Rx (Total)} & \textbf{Tx (User)} & \textbf{Rx (Total)} & \textbf{Tx (User)} & \textbf{Rx (Total)} \\ \hline
128 & 8 & 16 & 448 & 576 & 480 & 832 & 160 & 1408 & 192 & 1664 & \textbf{128} & \textbf{1664} \\ \hline
128 & 32 & 4 & 448 & 576 & 704 & 452 & 132 & 4352 & 136 & 4480 & \textbf{128} & \textbf{4480} \\ \hline
128 & 64 & 2 & 448 & 576 & 449 & 640 & 129 & 8384 & 130 & 8448 & \textbf{128} & \textbf{8448} \\ \hline
128 & 128 & 1 & 448 & 576 & 448 & 576 & 128 & 16512 & 128 & 16512 & \textbf{128} & \textbf{16512} \\ \hline
256 & 8 & 32 & 1024 & 1280 & 1920 & 1104 & 336 & 2944 & 416 & 3584 & \textbf{256} & \textbf{3584} \\ \hline
256 & 32 & 8 & 1024 & 1280 & 1036 & 1664 & 268 & 8832 & 280 & 9216 & \textbf{256} & \textbf{9216} \\ \hline
256 & 64 & 4 & 1024 & 1280 & 1028 & 1536 & 260 & 16896 & 264 & 17152 & \textbf{256} & \textbf{17152} \\ \hline
256 & 128 & 2 & 1024 & 1280 & 1025 & 1408 & 257 & 33152 & 258 & 33280 & \textbf{256} & \textbf{33280} \\ \hline
256 & 256 & 1 & 1024 & 1280 & 1280 & 1024 & 256 & 65792 & 256 & 65792 & \textbf{256} & \textbf{65792} \\ \hline
512 & 8 & 64 & 2304 & 2816 & 2496 & 4352 & 704 & 6144 & 896 & 7680 & \textbf{512} & \textbf{7680} \\ \hline
512 & 32 & 16 & 2304 & 2816 & 2336 & 3840 & 544 & 17920 & 576 & 18944 & \textbf{512} & \textbf{18944} \\ \hline
512 & 64 & 8 & 2304 & 2816 & 2316 & 3584 & 524 & 34048 & 536 & 34816 & \textbf{512} & \textbf{34816} \\ \hline
512 & 128 & 4 & 2304 & 2816 & 2308 & 3328 & 516 & 66560 & 520 & 67072 & \textbf{512} & \textbf{67072} \\ \hline
512 & 256 & 2 & 2304 & 2816 & 2305 & 3072 & 513 & 131840 & 514 & 132096 & \textbf{512} & \textbf{132096} \\ \hline
512 & 512 & 1 & 2304 & 2816 & 2304 & 2816 & 512 & 262656 & 512 & 262656 & \textbf{512} & \textbf{262656} \\ \hline
1024 & 8 & 128 & 5120 & 6144 & 9728 & 5568 & 1472 & 12800 & 1920 & 16384 & \textbf{1024} & \textbf{16384} \\ \hline
1024 & 32 & 32 & 5120 & 6144 & 8704 & 5200 & 1104 & 36352 & 1184 & 38912 & \textbf{1024} & \textbf{38912} \\ \hline
1024 & 64 & 16 & 5120 & 6144 & 8192 & 5152 & 1056 & 68608 & 1088 & 70656 & \textbf{1024} & \textbf{70656} \\ \hline
1024 & 128 & 8 & 5120 & 6144 & 5132 & 7680 & 1036 & 133632 & 1048 & 135168 & \textbf{1024} & \textbf{135168} \\ \hline
1024 & 256 & 4 & 5120 & 6144 & 5124 & 7168 & 1028 & 264192 & 1032 & 265216 & \textbf{1024} & \textbf{265216} \\ \hline
1024 & 512 & 2 & 5120 & 6144 & 5121 & 6656 & 1025 & 525824 & 1026 & 526336 & \textbf{1024} & \textbf{526336} \\ \hline
\end{tabular}%
}
\end{table*}

The reduced complexity stems from the simplified structure of P-OFDMA, which requires only an $M$-point IFFT/FFT at the user side (where $M = N/K$), compared to the $N$-point transforms plus additional DFT precoding required in SC-FDMA. For instance, with $N=256$ subcarriers and $K=64$ users, each user performs only a 4-point IFFT in P-OFDMA, whereas an SC-FDMA user must execute a 4-point FFT followed by a 256-point IFFT. As shown in Table~\ref{tab:complexity_comparison}, the uplink transmitters require 1028, 260, 264, and 256 complex multiplications for the SC-FDMA, P-OFDMA, P-OFDMA-DCT, and P-OFDMA-DFT methods, respectively. In the DFT-precoded version, only a multiplication by the diagonal matrix $\mathbf{X}_m$ is required, which makes the transmitter-side computational load very low.

\section{Performance Evaluation of P-OFDMA}\label{perform}

In this section, the performance of the proposed P-OFDMA method is analyzed through comprehensive comparisons with conventional OFDMA, SC-FDMA, and the precoded versions of the proposed P-OFDMA, namely P-OFDMA-DCT and P-OFDMA-DFT. The evaluation is conducted across key metrics, including Peak-to-Average Power Ratio (PAPR), Bit Error Rate (BER), and robustness to frequency-selective fading. Two modulation schemes, 16-QAM and 64-QAM, are examined to assess performance across different spectral efficiency levels.

\subsection{Simulation Setup}

All simulations were conducted with $N=256$ subcarriers and evaluated across multiple channel delay spread values (50, 100, 200, 300, 500, 750, 1000, 1500, 2000, 2500, 3000, and 3500~ns) to assess system robustness under varying frequency-selective fading conditions \cite{sreedhar2022refined}. The number of users was varied with $K \in \{16, 32, 64, 128\}$, corresponding to $N/K$ ratios of $\{16, 8, 4, 2\}$ subcarriers per user. Two modulation schemes, 16-QAM and 64-QAM, were examined, and the SNR was swept from 0 to 40~dB in 4~dB increments. Unless otherwise specified, the figures are obtained using a default configuration of 300~ns delay spread, $K=64$ users, and 16-QAM modulation, with the SNR set according to the metric being evaluated. The channel model assumes a tapped-delay-line multipath channel, where the average tap power decays exponentially with excess delay \cite{saleh1987}.

Each data point in the performance evaluation---including BER, average PAPR, and PAPR CCDF figures---is obtained through Monte Carlo simulations. The simulation framework is specifically designed to capture highly dynamic fading conditions. Instead of applying a static channel over a large block of data, a fresh and statistically independent multipath channel impulse response is generated for every transmitted OFDM block and for each individual user in the system. The transmitted signals propagate through these independent frequency-selective fading channels, followed by the addition of Additive White Gaussian Noise (AWGN) scaled to the target SNR level. At the receiver, the signals are equalized and demodulated. The resulting bit errors and PAPR samples are continuously accumulated over a sufficiently large number of transmitted blocks to ensure the statistical reliability and smoothness of the final averaged metrics and empirical distributions.

\begin{figure*}[!t]
	\centering
	\subfloat[$N/K=2$ PAPR Performance]{\includegraphics[width=0.48\textwidth]{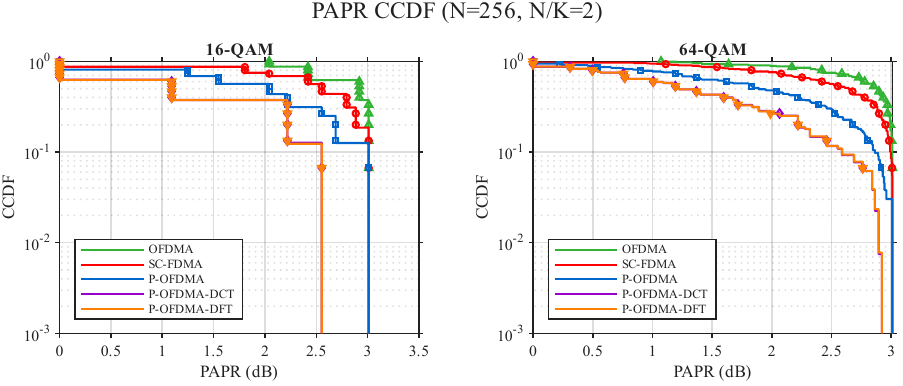}\label{fig:papr_2}}
	\hfil
	\subfloat[$N/K=4$ PAPR Performance]{\includegraphics[width=0.48\textwidth]{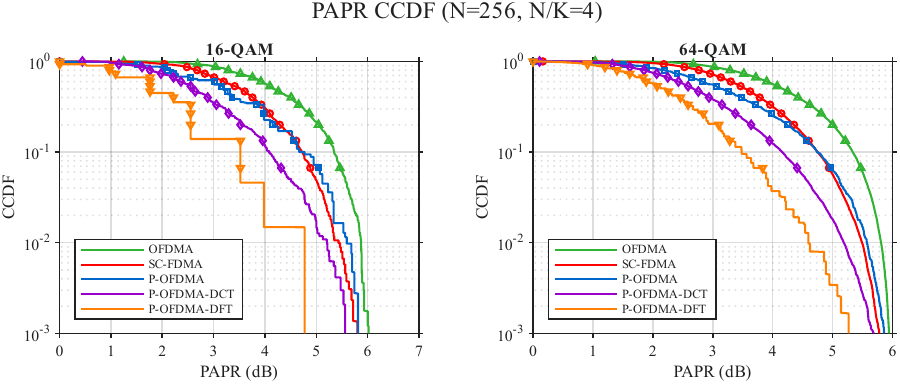}\label{fig:papr_4}}
	\\
	\subfloat[$N/K=8$ PAPR Performance]{\includegraphics[width=0.48\textwidth]{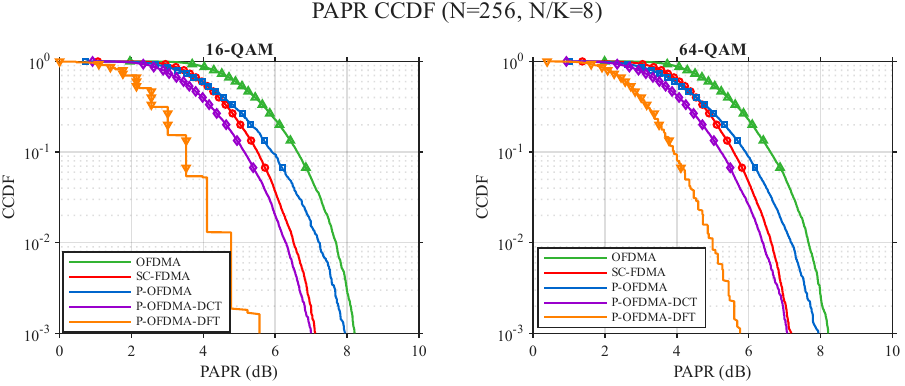}\label{fig:papr_8}}
	\hfil
	\subfloat[$N/K=16$ PAPR Performance]{\includegraphics[width=0.48\textwidth]{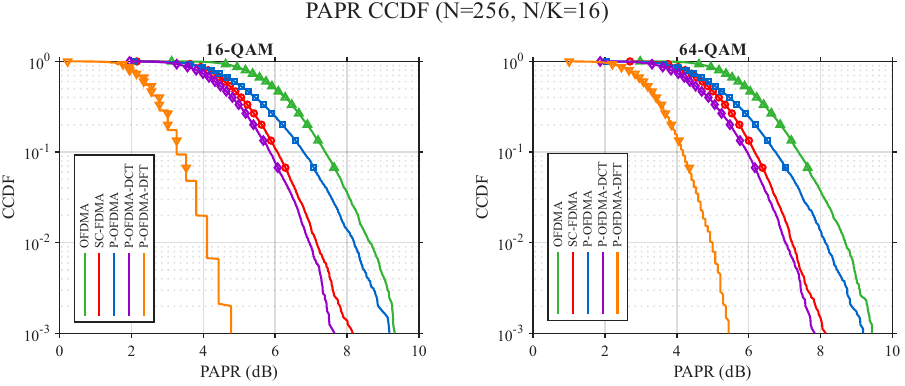}\label{fig:papr_16}}
	\caption{CCDF curves of PAPR performance for OFDMA, SC-FDMA, P-OFDMA, P-OFDMA-DCT, and P-OFDMA-DFT schemes at different $N/K$ ratios. Each empirical CCDF is computed from $500$ OFDM blocks $\times$ number of users per configuration over a 300~ns delay-spread channel. Columns from left to right represent 16-QAM and 64-QAM modulations, respectively.}
\label{fig:papr_karsilastirma}

\end{figure*}

\begin{figure*}[!t]
	\centering
	\includegraphics[width=0.95\textwidth,height=0.34\textheight,keepaspectratio]{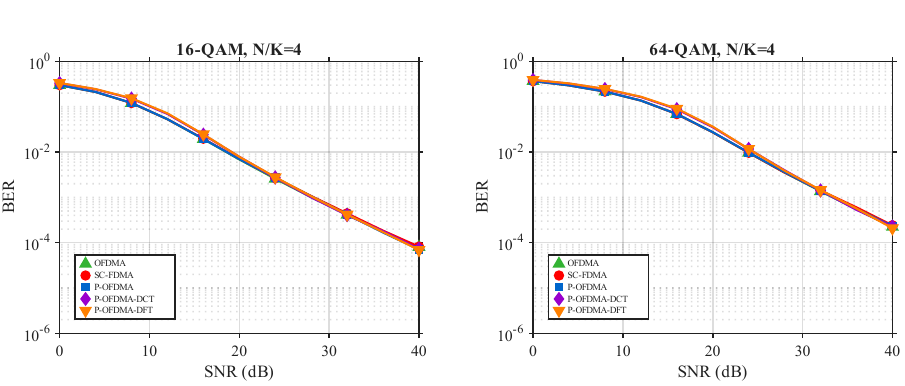}
	\caption{BER versus SNR performance comparison of OFDMA, SC-FDMA, P-OFDMA, P-OFDMA-DCT, and P-OFDMA-DFT for $N/K=4$ ($N=256$, $K=64$). Left: 16-QAM. Right: 64-QAM. Channel model uses 300~ns maximum delay spread.}
	\label{fig:ber_snr}
\end{figure*}

\begin{figure*}[!t]
	\centering
	\subfloat[16-QAM]{\includegraphics[width=0.42\textwidth]{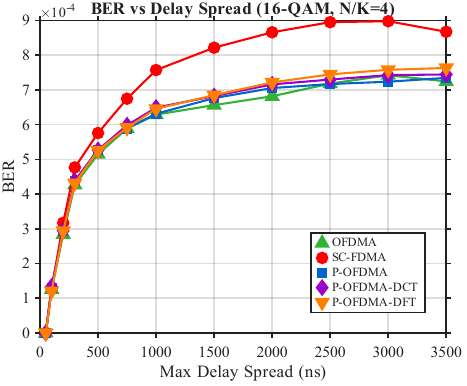}\label{fig:ber_delay_16qam}}
	\hfil
	\subfloat[64-QAM]{\includegraphics[width=0.42\textwidth]{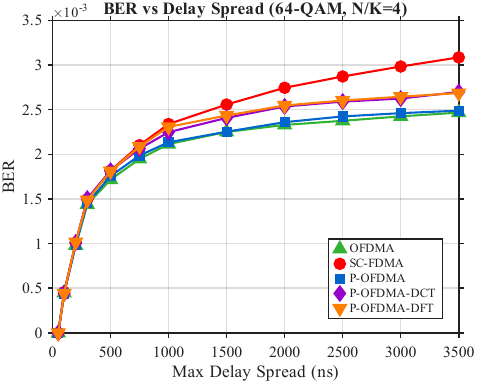}\label{fig:ber_delay_64qam}}
	\caption{BER versus maximum delay spread for OFDMA, SC-FDMA, P-OFDMA, P-OFDMA-DCT, and P-OFDMA-DFT schemes at $N/K=4$ ($N=256$, $K=64$) and 32~dB SNR. (a)~16-QAM modulation. (b)~64-QAM modulation.}
	\label{fig:ber_delay}
\end{figure*}

\subsection{PAPR Performance}

PAPR is a critical parameter in multi-carrier systems as it directly affects power amplifier efficiency. Figure \ref{fig:papr_karsilastirma} illustrates the PAPR performance of the five examined methods through complementary cumulative distribution function (CCDF) curves. A CCDF curve positioned further to the left indicates lower PAPR values and thus better performance. Each empirical CCDF curve in Fig.~\ref{fig:papr_karsilastirma} is computed from $500$ OFDM blocks $\times$ number of users per configuration.

The CCDF of PAPR is defined as
\begin{equation}
\mathrm{CCDF}(z) = \Pr\!\left(\mathrm{PAPR} > z\right).
\end{equation}

Across all $N/K$ scenarios, the P-OFDMA-DFT method clearly achieves the lowest PAPR values, consistently outperforming all other schemes including P-OFDMA-DCT. Conventional OFDMA exhibits the highest PAPR values, as expected. The comparison between P-OFDMA and SC-FDMA reveals interesting results depending on the $N/K$ ratio:

For $N/K=2$ (128 users sharing 256 subcarriers), the CCDF curves for 16-QAM show that P-OFDMA-DFT achieves the lowest PAPR, with the curve positioned furthest to the left and dropping below $10^{-1}$ around 2.5--2.6 dB. P-OFDMA and P-OFDMA-DCT follow closely around 3.0 dB and both outperform SC-FDMA, while OFDMA exhibits the worst performance. For 64-QAM at $N/K=2$, P-OFDMA-DCT performs similarly to P-OFDMA-DFT; both remain superior to P-OFDMA, SC-FDMA, and OFDMA.

For $N/K=4$ (64 users, 4 subcarriers per user), the performance gaps become more pronounced. For 16-QAM, P-OFDMA-DFT reaches $10^{-3}$ at approximately 4.7 dB, followed by P-OFDMA-DCT at about 5.6 dB. SC-FDMA and P-OFDMA show comparable performance, with CCDF values reaching $10^{-3}$ around 5.7--5.9 dB, while OFDMA remains the worst at approximately 6 dB. For 64-QAM, the same ordering is preserved.

For $N/K=8$, P-OFDMA-DFT maintains a clear advantage. In the 16-QAM case, it reaches $10^{-3}$ around 5.5 dB, whereas SC-FDMA and P-OFDMA-DCT reach $10^{-3}$ around 7 dB, and standard P-OFDMA together with OFDMA extends to about 8.0 dB.

For $N/K=16$, SC-FDMA begins to outperform standard P-OFDMA for 16-QAM, reaching $10^{-2}$ around 7.4 dB versus approximately 8.0 dB for P-OFDMA. However, P-OFDMA-DFT still outperforms all methods, reaching $10^{-2}$ around 4.0 dB, and P-OFDMA-DCT also remains better than SC-FDMA at approximately 6.9 dB.

Across modulation orders, higher-order modulation (64-QAM versus 16-QAM) generally increases PAPR for all schemes; however, P-OFDMA-DFT remains the most consistent and lowest-PAPR method, also benefiting from the fact that no addition operation is required in its transmitter-side processing. The crossover point at which SC-FDMA starts to outperform standard P-OFDMA occurs between $N/K=4$ and $N/K=8$ (approximately around $N/K=5$ depending on modulation), while this crossover does not apply to the precoded variants because P-OFDMA-DFT consistently outperforms SC-FDMA across all tested $N/K$ values.

\begin{figure}[!t]
	\centering
	\includegraphics[width=\columnwidth]{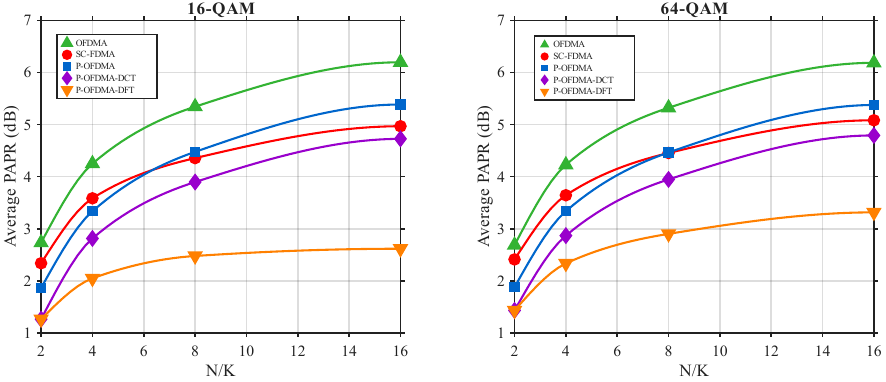}
	\caption{Average PAPR performance as a function of $N/K$ ratio for 16-QAM and 64-QAM modulation schemes. Each marker is computed by averaging over transmitted OFDM blocks generated under the Monte Carlo simulation framework over a 300~ns delay-spread channel.}

	\label{fig:analiz}
\end{figure}

This trend is confirmed by the average PAPR plots shown in Figure \ref{fig:analiz}.  The graphs demonstrate that P-OFDMA-DFT (orange downward triangles) consistently achieves the lowest average PAPR across all $N/K$ ratios, followed by P-OFDMA-DCT (purple diamonds). The P-OFDMA (blue squares) and SC-FDMA (red circles) curves intersect around $N/K=4$ to $5$, beyond which SC-FDMA becomes more advantageous than standard P-OFDMA. OFDMA (green upward triangles) maintains the highest average PAPR across all scenarios.

The quantitative analysis of average PAPR reveals several important trends:
\begin{itemize}
    \item \textbf{P-OFDMA-DFT Dominance:} P-OFDMA-DFT (orange line with downward triangles) maintains the lowest average PAPR across all $N/K$ ratios and both modulation schemes. For 16-QAM, it achieves approximately 1.2 dB at $N/K=2$, increasing to approximately 2.7 dB at $N/K=16$. For 64-QAM, the corresponding values are approximately 1.5 dB and 3.3 dB. This represents the best PAPR performance among all five methods tested, consistently outperforming even P-OFDMA-DCT by approximately 0.1--0.2 dB at low $N/K$ ratios and up to 2.0 dB at $N/K=16$.

    \item \textbf{P-OFDMA-DCT Performance:} P-OFDMA-DCT (purple dashed line with diamonds) achieves the second-lowest average PAPR. For 16-QAM, it ranges from approximately 1.3 dB at $N/K=2$ to 4.7 dB at $N/K=16$. While clearly outperformed by P-OFDMA-DFT, it still achieves substantially lower PAPR than SC-FDMA and standard P-OFDMA.

    \item \textbf{P-OFDMA vs. SC-FDMA Crossover:} For 16-QAM, at $N/K=2$, P-OFDMA achieves approximately 1.9 dB average PAPR compared to SC-FDMA's 2.4 dB, representing a 21\% improvement. The crossover occurs around $N/K=4$ to $5$, where both schemes achieve similar PAPR values of approximately 3.3--3.6 dB. Beyond this point, SC-FDMA shows lower PAPR than standard P-OFDMA, reaching approximately 5.0 dB at $N/K=16$ compared to P-OFDMA's 5.5 dB. For 64-QAM, the crossover behavior is similar, with P-OFDMA achieving approximately 1.9 dB versus SC-FDMA's 2.5 dB at $N/K=2$, and the crossover occurring around $N/K=5$.

    \item \textbf{OFDMA Performance:} Conventional OFDMA (green triangles) consistently exhibits the highest average PAPR values, starting from approximately 2.8 dB at $N/K=2$ and increasing to approximately 6.2 dB at $N/K=16$ for 16-QAM (and approximately 2.8 dB to 6.3 dB for 64-QAM). The steep slope of the OFDMA curve indicates its sensitivity to the number of subcarriers per user.
\end{itemize}

\subsection{Bit Error Rate (BER) Performance}

Figure \ref{fig:ber_snr} shows the BER versus SNR performance for $N/K=4$ ($K=64$ users) at 300~ns delay spread for both 16-QAM and 64-QAM modulations. All five schemes---OFDMA, SC-FDMA, P-OFDMA, P-OFDMA-DCT, and P-OFDMA-DFT---achieve nearly identical BER performance across the entire SNR range from 0 to 40 dB. This is an important finding, as it demonstrates that the significant PAPR advantages offered by the precoded P-OFDMA variants come without any BER penalty.

For 16-QAM, all schemes converge tightly, with BER decreasing from approximately $5 \times 10^{-1}$ at 0 dB SNR to approximately $8 \times 10^{-5}$ at 40 dB SNR. For 64-QAM, the BER curves again overlap closely, ranging from approximately $6 \times 10^{-1}$ at 0 dB to approximately $2 \times 10^{-4}$ at 40 dB. The virtually indistinguishable BER performance confirms that the proposed P-OFDMA schemes achieve their PAPR reduction without sacrificing error performance.

\subsection{Frequency Selectivity Robustness}

Figure \ref{fig:ber_delay} presents the BER versus maximum delay spread characteristics for all five schemes at $N/K=4$, $K=64$, and 32~dB SNR, evaluated from 50 ns to 3500 ns for both 16-QAM and 64-QAM.

The BER versus delay spread analysis reveals critical differences in how the schemes handle frequency-selective fading:
\begin{itemize}
	\item \textbf{SC-FDMA Vulnerability:} SC-FDMA (red circles) shows dramatically higher BER than all other schemes across most delay spread values. This behavior is consistent with the well-known property that the DFT precoding in SC-FDMA spreads each modulation symbol across all of a user's allocated subcarriers, so a single deeply faded subcarrier degrades all the symbols carried in that block~\cite{ciochina2007analysis}. For 16-QAM, SC-FDMA's BER increases from near zero at 50 ns to approximately $9.0 \times 10^{-4}$ at 2500--3000 ns before slightly decreasing at 3500 ns. This is significantly worse than the other four methods, which cluster together at much lower BER values. For 64-QAM, the same trend is observed: SC-FDMA reaches approximately $3.0 \times 10^{-3}$ at high delay spreads, while the other schemes remain around $2.5 \times 10^{-3}$ or below.

    \item \textbf{OFDMA and P-OFDMA Robustness:} OFDMA (green triangles) and P-OFDMA (blue squares) demonstrate the best robustness to frequency-selective fading, achieving the lowest BER values across most delay spread conditions. For 16-QAM, both schemes maintain BER below approximately $7.3 \times 10^{-4}$ even at 3500 ns delay spread. Their distributed subcarrier allocation across the entire frequency band enables effective frequency diversity exploitation, a property that has previously been shown to substantially improve the performance of OFDMA-family schemes in non-frequency-adaptive uplink scenarios~\cite{svensson2009bifdma}.

    \item \textbf{Precoded Variants (DCT and DFT):} P-OFDMA-DCT (purple diamonds) and P-OFDMA-DFT (orange downward triangles) exhibit slightly higher BER than standard P-OFDMA and OFDMA at high delay spreads, particularly above 1000 ns. For 16-QAM, P-OFDMA-DFT reaches approximately $7.5 \times 10^{-4}$ at 3000--3500 ns, while P-OFDMA-DCT shows similar values around $7.3$--$7.5 \times 10^{-4}$. This slight BER increase relative to standard P-OFDMA (approximately $7.2 \times 10^{-4}$) is marginal and represents a negligible trade-off compared to the substantial PAPR gains.

	\item \textbf{Practical Implications:} For 16-QAM, the BER difference between SC-FDMA and the P-OFDMA variants is most pronounced at delay spreads between 500 ns and 3000 ns. At 1000 ns delay spread, SC-FDMA achieves approximately $7.5 \times 10^{-4}$ while P-OFDMA achieves approximately $6.3 \times 10^{-4}$, representing a 16\% improvement. At 3000 ns, the gap is even larger: SC-FDMA at $9.0 \times 10^{-4}$ versus P-OFDMA at $7.2 \times 10^{-4}$, a 20\% improvement. For 64-QAM, SC-FDMA's BER is approximately $3.0 \times 10^{-3}$ at 3500 ns versus approximately $2.5 \times 10^{-3}$ for P-OFDMA, representing a 17\% improvement. The P-OFDMA-based schemes' superior robustness to frequency-selective fading makes them particularly suitable for deployment in urban and suburban environments.

\end{itemize}

\subsection{Power Spectral Density}

\begin{figure*}[!t]
	\centering
	\subfloat[User 1 PSD]{\includegraphics[width=0.31\textwidth]{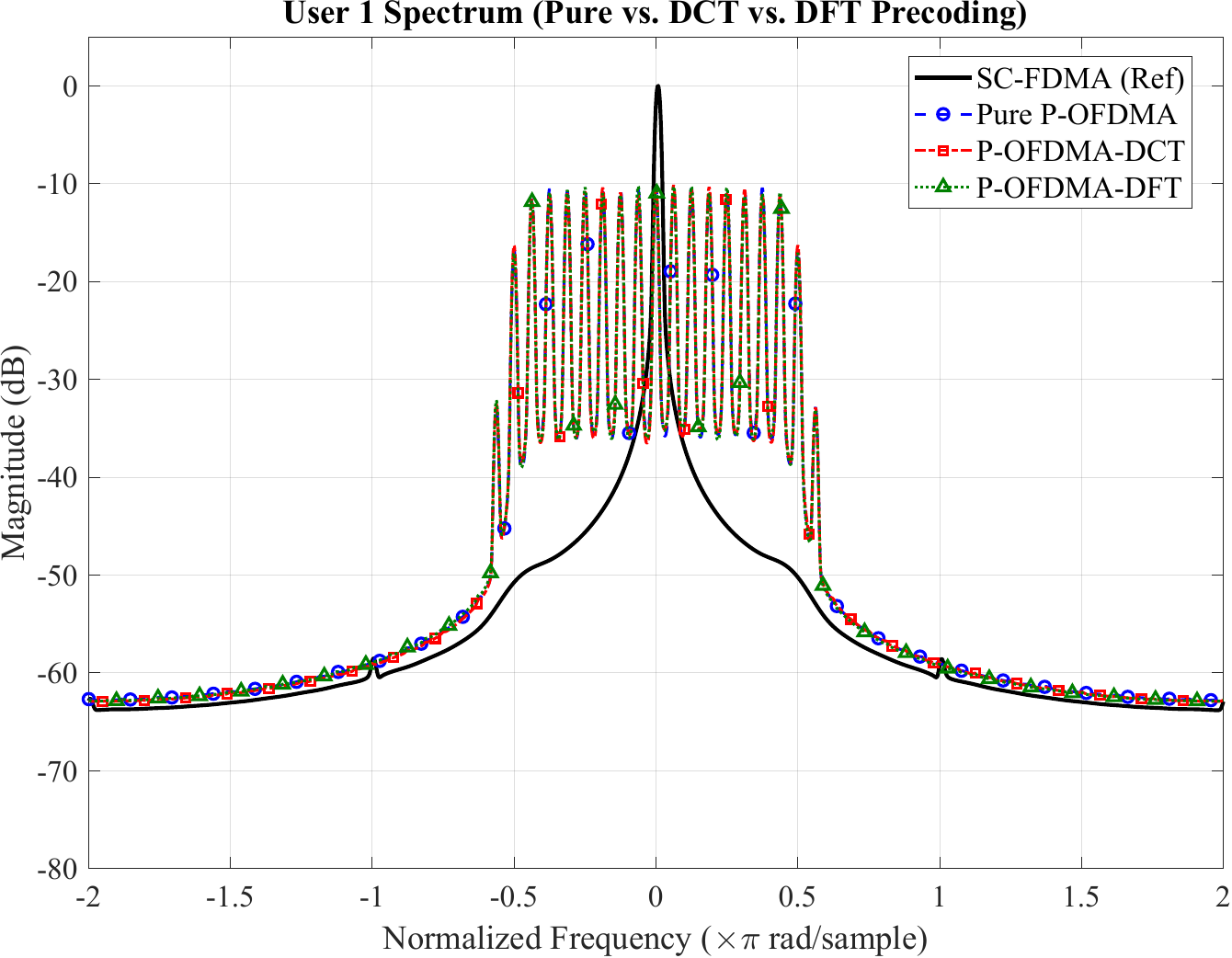}\label{fig:psd_user1}}
	\hfil
	\subfloat[User 32 PSD]{\includegraphics[width=0.31\textwidth]{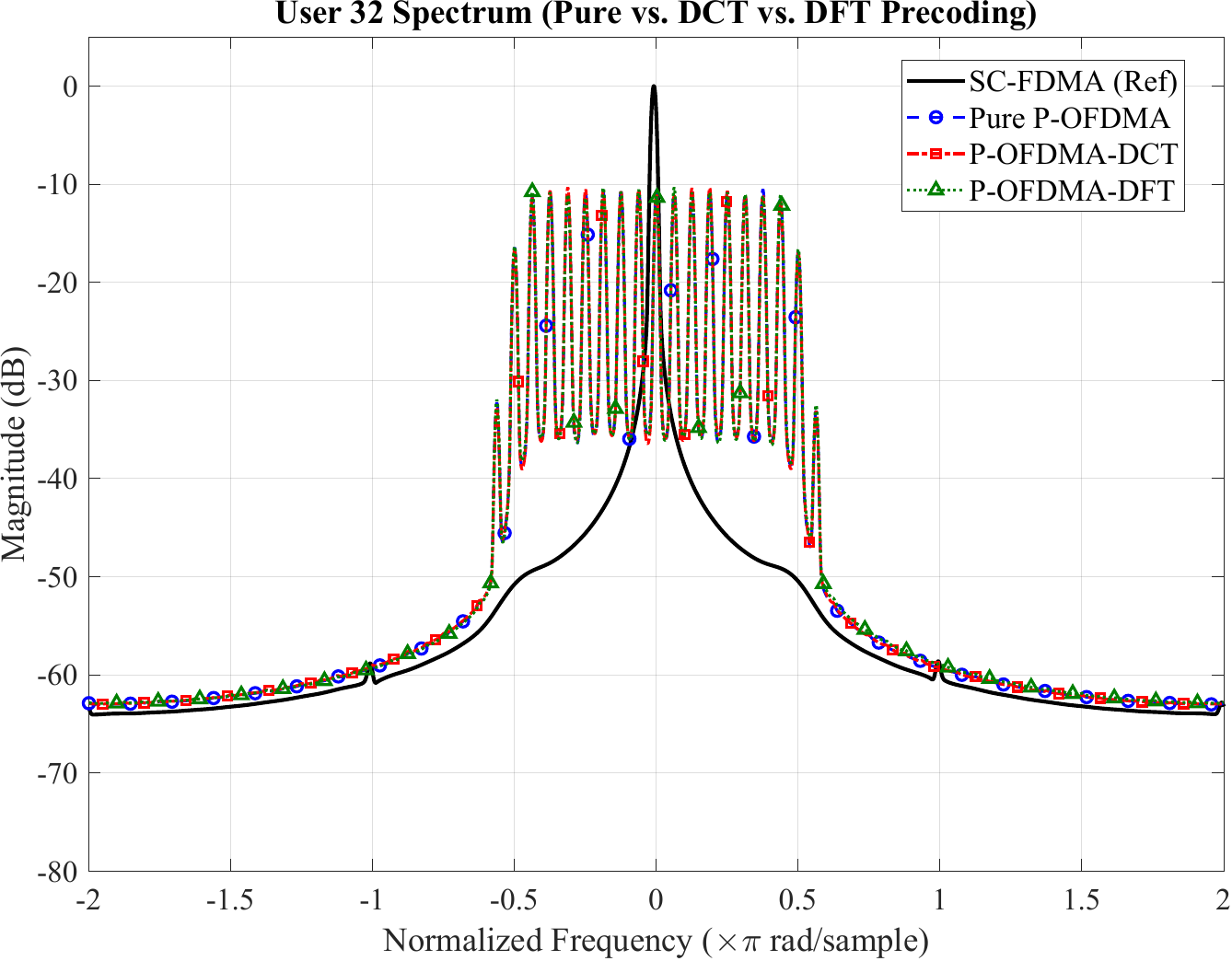}\label{fig:psd_user32}}
	\hfil
	\subfloat[Total Band PSD]{\raisebox{1.2mm}{\includegraphics[width=0.31\textwidth]{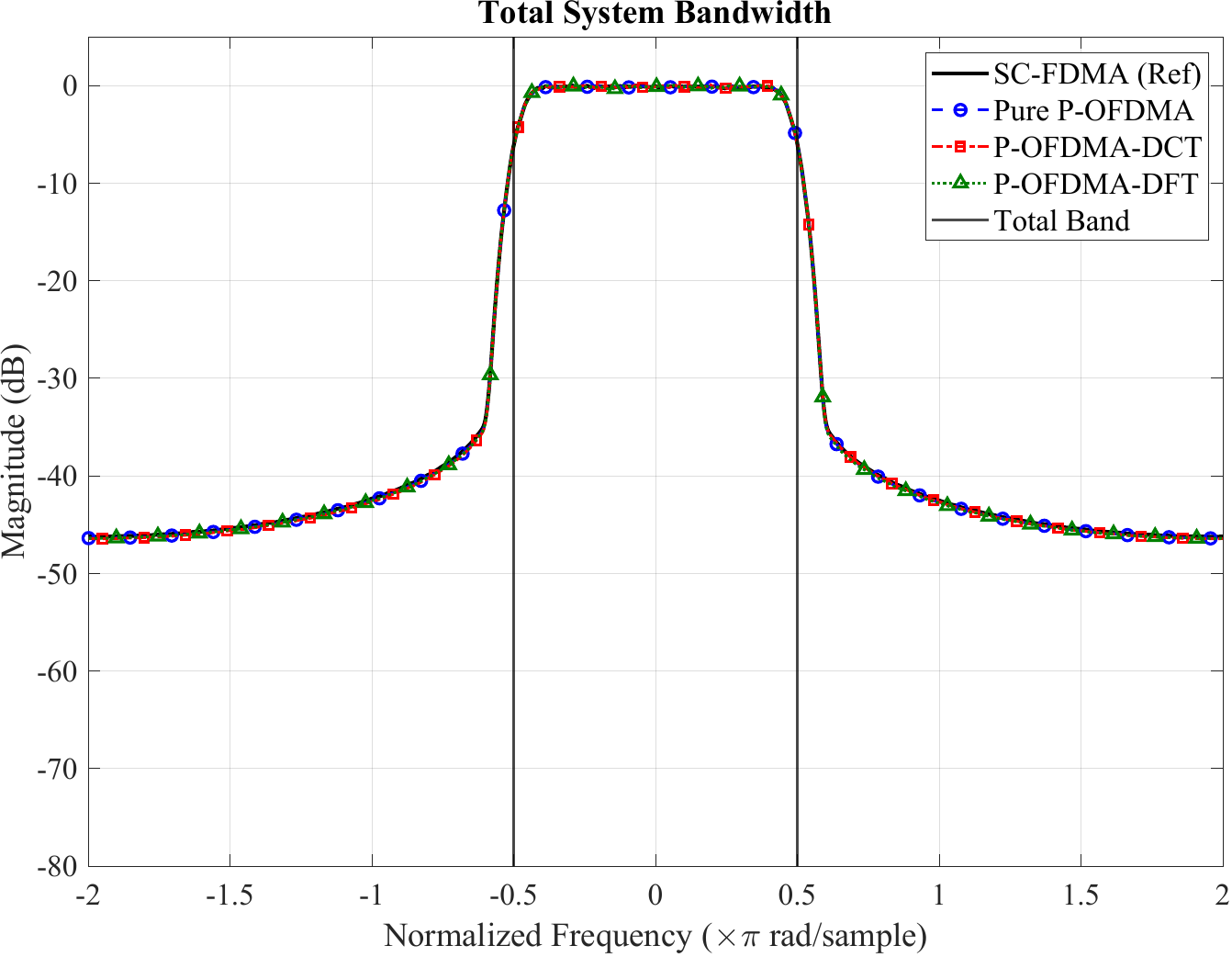}}\label{fig:psd_totalband}}
	\caption{Welch-based power spectral density (PSD) results for the P-OFDMA uplink configuration with $N=1024$ subcarriers and $K=64$ users: (a) user 1 spectrum, (b) user 32 spectrum, and (c) aggregate occupied-band spectrum. The PSD estimates are obtained via Welch's method and illustrate the periodic allocation structure together with the overall band occupancy.}
	\label{fig:psd_karsilastirma}
\end{figure*}

Figure \ref{fig:psd_karsilastirma} presents the PSD outputs obtained from the $N=1024$, $K=64$ P-OFDMA simulation using Welch's method, including representative spectra for user 1 (Fig. \ref{fig:psd_user1}), user 32 (Fig. \ref{fig:psd_user32}), and the aggregate occupied band (Fig. \ref{fig:psd_totalband}). As clearly seen in Figs. \ref{fig:psd_user1}--\ref{fig:psd_totalband}, the PSD interpretation is based directly on these three subplots.

As observed in Fig. \ref{fig:psd_user1} and Fig. \ref{fig:psd_user32}, both user 1 and user 32 utilize periodically spaced tones over the entire frequency range rather than a localized contiguous portion, confirming that P-OFDMA effectively exploits full-band frequency diversity even at the individual-user level. The aggregate (total-band) PSD further indicates that P-OFDMA does not enlarge the occupied transmission bandwidth.

In addition, the out-of-band behavior remains well controlled: the sideband leakage profile is consistent with the SC-FDMA benchmark, indicating comparable spectral containment performance. This is a practically important result because it shows that the PAPR advantages of P-OFDMA are achieved without sacrificing adjacent-band emission characteristics or spectral mask compatibility. Overall, the PSD analysis supports that periodic mapping provides full-band access per user while preserving bandwidth efficiency and leakage performance.

\section{Conclusion}

In this letter, a novel multiple access scheme, Periodic Orthogonal Frequency Division Multiple Access (P-OFDMA), was proposed to address the high peak-to-average power ratio (PAPR) challenge in uplink OFDMA systems. Furthermore, two precoded versions, P-OFDMA-DCT and P-OFDMA-DFT, were introduced to further enhance PAPR performance while maintaining low computational complexity at the transmitter side. This low processing requirement makes the proposed schemes particularly suitable for uplink transmitters in mMTC/IoT systems, where devices typically have limited computational capability. A comprehensive performance evaluation was conducted, comparing the proposed schemes against conventional OFDMA and SC-FDMA across key metrics including PAPR, Bit Error Rate (BER), frequency selectivity robustness, power spectral density, and computational complexity.

The results unequivocally demonstrate the significant advantages of the proposed methods. P-OFDMA-DFT emerged as the superior scheme across all tested scenarios, consistently achieving the lowest PAPR among all five methods. At $N/K=2$ with 16-QAM, P-OFDMA-DFT achieves an average PAPR of approximately 1.2 dB, and remains as low as approximately 2.7 dB even at $N/K=16$. This represents a dramatic improvement over conventional OFDMA (which reaches 6.2 dB at $N/K=16$) and SC-FDMA (approximately 5.0 dB at $N/K=16$). P-OFDMA-DCT also achieves lower PAPR than SC-FDMA and OFDMA. Crucially, P-OFDMA-DFT achieves this PAPR advantage while also having the lowest computational complexity at the transmitter, requiring only $N$ complex multiplications with zero additions. Although the receiver-side computational load is higher than that of OFDMA and SC-FDMA, the total processing load considering all uplink P-OFDMA users together with the receiver remains lower overall.

A key finding is that all five schemes achieve virtually identical BER performance across the entire SNR range (0--40 dB) for both 16-QAM and 64-QAM modulations, confirming that the PAPR reduction offered by the P-OFDMA variants comes at no cost to error performance. This is a critical advantage for practical deployment, as it means that system designers can obtain substantial PAPR gains without any BER trade-off.

The BER versus delay spread analysis revealed that SC-FDMA is significantly more vulnerable to frequency-selective fading than all P-OFDMA variants. For 16-QAM at $N/K=4$, SC-FDMA's BER reaches approximately $9.0 \times 10^{-4}$ at high delay spreads, while all P-OFDMA-based schemes maintain BER values around $7.2$--$7.5 \times 10^{-4}$, representing a 17--20\% improvement. OFDMA and P-OFDMA show the best frequency selectivity robustness, while the precoded variants (DCT and DFT) exhibit only marginally higher BER at very high delay spreads, confirming that the precoding does not significantly degrade robustness.

The proposed P-OFDMA scheme proved to be a highly effective and balanced solution. For scenarios with a low number of subcarriers per user ($N/K \leq 4$), standard P-OFDMA achieved substantial PAPR advantages over SC-FDMA, with approximately 21\% improvement at $N/K=2$ for 16-QAM. The critical crossover point where SC-FDMA begins to outperform standard P-OFDMA in PAPR occurs between $N/K=4$ and $N/K=5$, providing clear guidance for system designers. However, this crossover is irrelevant for the precoded variants: P-OFDMA-DFT outperforms SC-FDMA in PAPR across all $N/K$ ratios tested.

The PSD analysis confirmed that all P-OFDMA variants maintain identical spectral characteristics to OFDMA and SC-FDMA, with no degradation in in-band or out-of-band performance. This ensures full compatibility with existing spectral mask requirements.

The periodic subcarrier allocation pattern of P-OFDMA enables all users to access the entire frequency band, maximizing frequency diversity gain and simplifying resource allocation procedures. This characteristic is particularly valuable for mMTC scenarios in 5G and beyond, where a large number of devices require sporadic access to small spectrum resources with limited battery capacity.

In conclusion, the proposed schemes present compelling alternatives to existing multiple access techniques. P-OFDMA-DFT stands out as the optimal choice, offering the lowest PAPR, the lowest transmitter computational complexity, and identical BER performance. P-OFDMA-DCT provides the second-best PAPR performance. Standard P-OFDMA offers excellent frequency diversity and lower PAPR than SC-FDMA for $N/K \leq 4$. These results establish the P-OFDMA family as a powerful and practical solution set for next-generation wireless communication systems, particularly for uplink transmissions in massive IoT and mMTC applications where devices have limited subcarrier allocations, constrained computational capability, and limited energy resources.

Future work will focus on extending P-OFDMA to support variable user loads and dynamic resource allocation strategies that can adaptively select between P-OFDMA variants based on the instantaneous $N/K$ ratio . Additionally, investigating its performance under imperfect channel state information and synchronization errors, and the integration of P-OFDMA with advanced techniques such as non-orthogonal multiple access (NOMA) warrant further investigation. Optimization of periodic allocation patterns for specific deployment scenarios, multi-cell coordination schemes to manage inter-cell interference, and the development of adaptive modulation and coding schemes specifically tailored for P-OFDMA also present promising research directions.

\section*{Acknowledgments}
This should be a simple paragraph before the References to thank those individuals and institutions who have supported your work on this article.

\begingroup
\emergencystretch=2em
\printbibliography
\endgroup

\vfill
\end{document}